\documentclass[glov3,twocolumn]{svjour3}
\usepackage{hyperref}
\usepackage{textcomp}
\usepackage{siunitx}
\usepackage{upgreek}
\usepackage{amsmath,amssymb}
\usepackage{graphicx,color}
\usepackage{microtype}
\usepackage[english]{babel}

\newcommand\thickrule[1][5mm]{\rule[0.4ex]{#1}{2pt}}

\newcommand\thickdashedrule[1][1mm]{\protect{\mbox{%
        \thickrule[#1]\hspace{#1}\thickrule[#1]\hspace{#1}\thickrule[#1]}}}

\definecolor{c4}{rgb}{0.2,0.6274,0.1725}
\definecolor{orange}{rgb}{1,0.5,0}

\newcommand{\blacksolid}{\textcolor{black}{\thickrule}}

\newcommand{\blackdashed}{\textcolor{black}{\thickdashedrule}}

\newcommand{\bra}[1]{\langle#1|}
\newcommand{\bral}[1]{\langle#1}
\newcommand{\ket}[1]{|#1\rangle}

\begin{document}


\title{A folded-sandwich polarization-entangled two-color photon pair source
with large tuning capability for applications in hybrid quantum architectures}

\author{Otto Dietz \and Chris M\"uller\and Thomas Krei{\ss}l \and Ulrike Herzog
\and \\
Tim Kroh\and Andreas Ahlrichs\and  Oliver Benson}

\institute{Nano-Optics Group, Institut f\"{u}r Physik, Humboldt-Universit\"{a}t
zu Berlin, Germany \email{otto.dietz@physik.hu-berlin.de}} 



\maketitle

\begin{abstract}
We demonstrate a two-color entangled photon pair source which can be adapted
easily to a wide range of wavelengths combinations. A Fresnel rhomb as a geometrical
quarter-wave plate and a versatile combination of compensation crystals are key components of the source. 
Entanglement of two photons at the Cs D1 line (894.3\,nm) 
and at the telecom O-band (1313.1\,nm) with a fidelity of $F=0.753 \pm 0.021$ 
is demonstrated and improvements of the setup are discussed. 
\end{abstract}

\keywords{down-conversion -- non-degenerate -- two-color -- entangled photon source -- quantum hybrid -- 
quantum repeater} 


\section{Introduction}

In recent years there has been an increasing effort to realize and study quantum hybrid systems. These consist of two dissimilar systems which are in a joint quantum state. 
Aside from the fundamental insight gained from studying such a peculiar, perhaps multi-particle entangled state, there are also immediate applications in quantum information processing.
Entanglement of a stationary and flying qubit, i.e., an electronic state with a long coherence time and a photon, respectively, represents a coherent quantum interface. 
Such interfaces are mandatory components of a quantum repeater
\cite{briegel_quantum_1998}, where entanglement has to be established between distant nodes. 
Experimental realizations demonstrated entanglement between a photon and
a stationary state in atoms \cite{volz_observation_2006}, ions
\cite{blinov_observation_2004}, semiconductor quantum
dots \cite{gao_observation_2012}, color defect centers
\cite{togan_quantum_2010}
and even superconducting qubits \cite{wallraff_strong_2004}. A Bell measurement
on two photons of each of two such states would then immediately establish
entanglement between two stationary, possibly dissimilar states
\cite{hofmann_heralded_2012,bernien_heralded_2013}.  

\begin{figure}
        \centering\includegraphics[width=0.7\columnwidth]{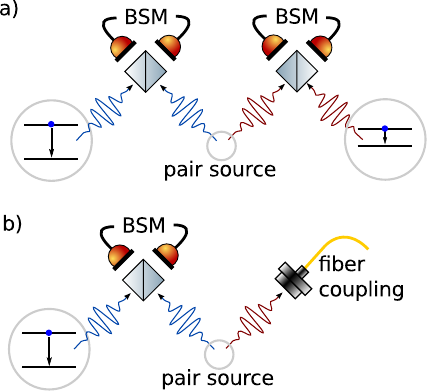}
\caption{Scheme of entanglement distribution via a two-color entangled photon source.  
		a) Two Bell-state measurements (BSM) on photons from two stationary
		qubit/flying qubit entangled systems and two photons from a two-color entangled photon source establish entanglement between two stationary qubits.
		b) Similarly, a single Bell-state measurement on a photon from a stationary
		qubit/flying qubit entangled system and one photon from a two-color entangled photon source establishes entanglement between the stationary qubit and a telecom photon. 
		\label{fig:schematic_entangle}}
\end{figure}

However, an intrinsic problem of entangling two dissimilar systems via photons in such a way is that there are system-specific transitions in the individual stationary systems. 
At the same time, if the entanglement should be established over long distances, e.g. via optical fibers, the photons have to be in the telecom band. 
One possible solution of this problem is to use a pair of entangled photons of
different wavelengths, which are matched to the two transitions of the
stationary systems or to one transition and the telecom band (see
Fig.~\ref{fig:schematic_entangle}). 

\begin{figure}
        \centering\includegraphics[width=\columnwidth]{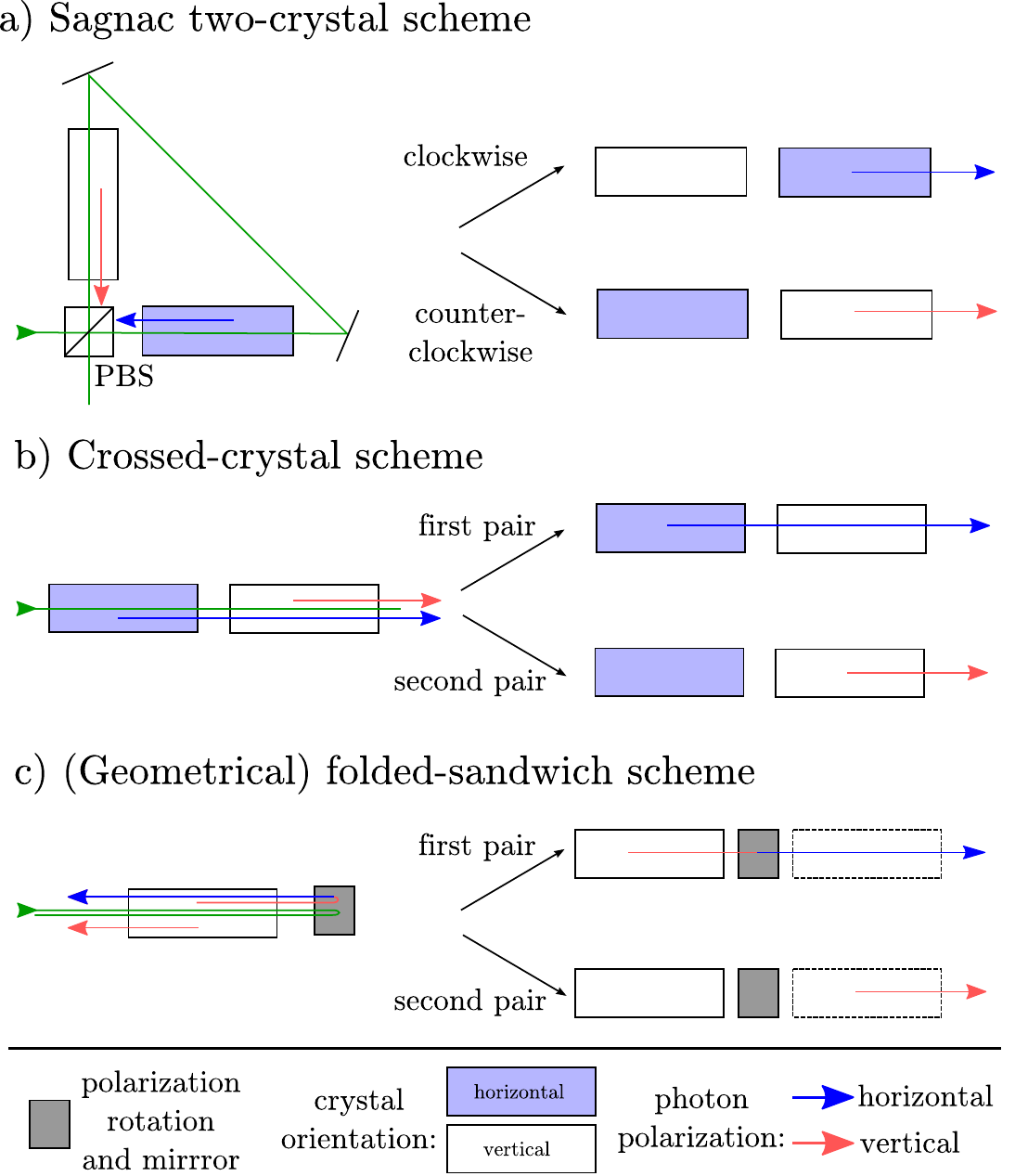}
\caption{Different schemes of two-color collinear entangled photon pair sources
		  using type-0 down-conversion. A horizontally
		  (vertically) polarized pump photon in an appropriately, i.e.,
		  horizontal (vertical) crystal
		  creates horizontally (vertically) polarized signal and idler photons.
		  No photons are created when pump polarization and crystal orientation
are orthogonal.
	a)~The two possible paths in the two-crystal Sagnac
        configuration. In the clockwise and counter-clockwise path the photon pair is generated in the second crystal. In
        the (counter) clockwise path, the photon pair is generated 
        vertically (horizontally) because the second crystal has horizontal
		  (vertical) orientation.
        b)~The two paths in the crossed-crystal configuration. The first photon pair
        (vertically polarized) is created in the first crystal and accumulates
        an extra phase while passing through the second crystal. The second
        photon pair is created in the last crystal, just as the counter-clockwise photon in the
        Sagnac configuration. 
        c)~Folded-sandwich configuration. The polarization rotation element and
		  a mirror (both depicted as one gray box) rotate and reflect the first pair. This is equivalent to the first pair in the
		  crossed-crystal scheme, only that instead of the crystal the light
		  polarization is
		  rotated. In the folded-sandwich configuration the second crystal is the
		  mirrored first crystal. 
        The (diagonal or unpolarized) pump beam is depicted in green. 
		  \label{fig:scheme}}
\end{figure}

Up to now the brightest and most practical sources of entangled photon pairs
rely on spontaneous parametric down-conversion in non-linear crystals
\cite{kwiat_new_1995}. With appropriate phase-matching 
two-color sources with photons at different wavelengths can be realized
\cite{pelton_bright_2004,hentschel_three-color_2009,stuart_flexible_2013}. For
subsequent fiber coupling collinear down-conversion schemes are most convenient. 
Furthermore, the collinear alignment enables further integration and can be used
in a monolithic design \cite{hentschel_three-color_2009}. 
There are different possible collinear arrangements. Figure~\ref{fig:scheme} shows the two-crystal Sagnac \cite{stuart_flexible_2013}, the
crossed-crystal \cite{trojek_collinear_2008}, and the
folded-sandwich configuration \cite{steinlechner_phase-stable_2013}.

The rotation of the second crystal in the crossed-crystal configuration (Fig.~\ref{fig:scheme} b) can be replaced by a half wave
plate. Replacing the half wave plate and the second crystal by a quarter-wave plate and a mirror yields
the folded-sandwich scheme (Fig.~\ref{fig:scheme} c).  
Using only a single non-linear crystal makes it considerably easier to avoid the leakage of which-crystal information.
The folded-sandwich and the crossed-crystal configuration include a compensation crystal, which compensates the
additional dispersion, whereas this is not needed in a two-crystal Sagnac
configuration.

Together with the pump, any two-color entangled photon source involves three
fields of widely different wavelengths.
This imposes strong constraints on dispersion compensation. In particular, a source for generating entanglement in quantum hybrid systems should be easy to align, intrinsically stable, and 
tunable in order to account for various transition frequencies.    
The Sagnac configuration requires a special three-color beam splitter and is very difficulty to align \cite{steinlechner_phase-stable_2013}.
The crossed-crystal configuration lacks the phase stability of the Sagnac
configuration.
Therefore it is desirable to use the simpler folded-sandwich configuration. In
this paper we describe the realization of such a configuration. In contrast to
previous work, we employ geometrical, i.e., wavelength independent, polarization manipulation. 

Here, we target for a two-color entangled photon source at $\lambda_s=894.3$\,nm
and $\lambda_i=1313.1$\,nm. These two wavelengths correspond to the Cs D1 line
and the telecom O-band, respectively. The former has been chosen on the one hand as
a convenient standard atomic transition. On the other hand it is also accessible
with excitonic transitions in InGaAs quantum dots \cite{ding_tuning_2010}. 
The source is thus applicable for quantum hybrid architectures
\cite{akopian_hybrid_2011,siyushev_molecular_2014} involving atomic species,
semiconductor quantum dots or molecules as well as long distance transfer via optical fibers. 
\section{Setup}
\begin{figure}
\centering\includegraphics[width=\columnwidth]{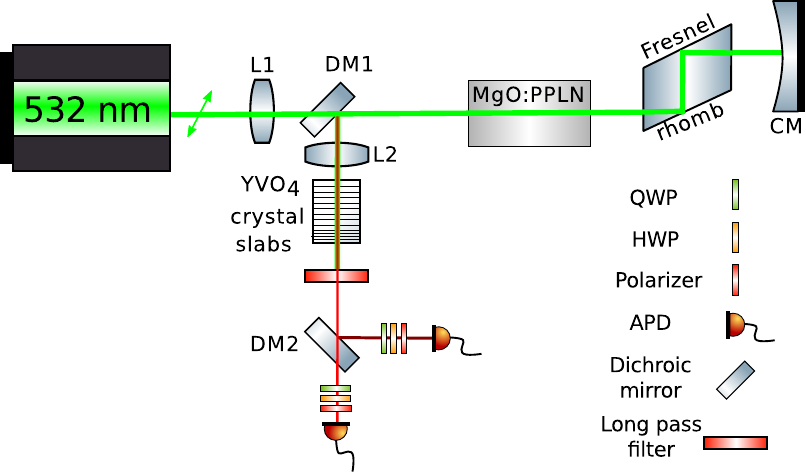}
\caption{  
        Setup based on one periodically poled lithium niobate crystal (ppLN) doped
        with approximately 5\% magnesium oxide (MgO), of length $L=40$\,mm. The crystal temperature is
        controlled via a crystal oven (not shown). The Fresnel rhomb acts as a geometrical quarter-wave
        plate. The concave mirror (CM) reflects the light back into the crystal.
        The focusing (L1) and collimation (L2) lenses adjust the beam
        width. The first dichroic mirror (DM1) separates pump and down-converted photons.
        The second dichroic mirror (DM2) separates signal and idler photons.
        The phase compensation crystal consists of several YVO$_4$ slabs to
        allow for dispersion control for a broad range of wavelengths. 
		  Both crystal ovens are omitted in this figure \cite{alexander_franzen_componentlibrary:_????}.
        \label{fig:setup}}
\end{figure}

The setup is shown in Fig.~\ref{fig:setup}. It resembles a folded-sandwich, but
the specifically tailored achromatic quarter-wave plate which is used in \cite{steinlechner_phase-stable_2013} is replaced by a wavelength independent Fresnel rhomb. 
The Fresnel rhomb yields a $\lambda/4$ phase shift after two internal reflections.
A diagonal polarized pump laser (532\,nm cw) is
focused into the non-linear crystal (length 40\,mm, facet 4\,mm$\times$1\,mm, type-0
phase matching, grating period of
7 $\upmu$m, multiple gratings,  HC Photonics Corp.). 
In this first pass, the vertical component of the diagonal pump beam may create a
vertically polarized pair. This pair and the pump beam propagate through the Fresnel rhomb.
The Fresnel rhomb is oriented at 45$^\circ$, such that it acts on the vertically polarized
photons as a quarter-wave plate. The vertical pair leaves the rhomb circularly
polarized.  The diagonal pump beam passes the Fresnel rhomb without modification.
The concave mirror (CM) reflects the light back onto the Fresnel rhomb and into
the crystal. The mirror is adjusted to reflect the light into the
same spatial mode. 
On the way back the Fresnel rhomb rotates the circularly polarized pair into
a horizontally polarized pair. 

The diagonal pump beam and the horizontally polarized pair pass the
non-linear crystal a second time. 
Again, the vertical part of the diagonal pump beam may create a vertically
polarized pair. The pump power is chosen such that the probability for creating more
than one pair per double-pass is low.
The pump beam and the created photons are separated at the
dichroic mirror (DM1). The photon pair is then collimated (L2) and directed onto a compensation crystal.
Finally, signal and idler photons are separated at a dichroic
mirror (DM2). In each of the separate arms, wave-plates and polarizers allow
for measurement of different polarization states. Each arm is coupled to a
single-mode fiber. The fibers can either be connected to a spectrograph or to avalanche photo diodes (APDs) for spectral or coincidence measurements.

In order to generate an entangled state of the form
\begin{equation}
		  \ket{\psi} = \frac{1}{\sqrt{2}}\left(
		  \ket{VV}+e^{i\phi}\ket{HH}\right).
		  \label{}
\end{equation}
it is required to compensate for the different dispersive characteristics
accumulated by the first pair
due to the extra pass through the rhomb and the non-linear crystal. Otherwise
this accumulated which-crystal information
would diminish the entanglement.

The extra phase $\phi$ of the first pair $\ket{HH}$ is given by
\begin{align}
		    \phi(\lambda_s,\lambda_i,T) = &\left(  \frac{n_o(\lambda_s,T)}{\lambda_s} +\frac{n_o(\lambda_i,T)}{\lambda_i}\right) 
		  		L_\text{crystal}\\ 
				&+
			 \left(  \frac{n_r(\lambda_s)}{\lambda_s}
			 +\frac{n_r(\lambda_i)}{\lambda_i}\right)
				2\,L_\text{rhomb},
		  \label{}
\end{align}
where $n_r$ is the refractive index of the Fresnel rhomb (BK7 glass of length
$L_\text{rhomb}$) and $n_o$
is the refractive index of the ordinary (horizontal) non-linear crystal axis of
length $L_\text{crystal}$. The refractive
index of the non-linear crystal depends on the crystal temperature $T$ and the
wavelengths of the down-converted signal and idler photons $\lambda_{s/i}$.

In our experiment, we chose undoped YVO$_4$ as a birefringent compensation crystal. A crystal of length  $\tilde L$ adds a phase of
\begin{align*}
		    \tilde \phi(\lambda_s,\lambda_i,T) = &\left(
			 \frac{\tilde n_o(\lambda_s,T)}{\lambda_s} +\frac{\tilde
			 n_o(\lambda_i,T)}{\lambda_i}\right.\\
			 &\left.- \frac{\tilde n_e(\lambda_s,T)}{\lambda_s}
			 -\frac{\tilde n_e(\lambda_i,T)}{\lambda_i} \right)
		  		\tilde L,
\end{align*}
where $\tilde n_{o/e}$ is the refractive index of the ordinary and extra-ordinary
axis of YVO$_4$.

The flat phase condition for optimum compensation
with the two different photon wavelengths $\lambda_s$ and $\lambda_i$ reads:
\begin{align}
		  \frac{d}{d\lambda}\left( \phi + \tilde \phi
		  \right)\Big|_{\lambda_{s,i}}&=0.
\end{align}

There are different values for the birefringent properties of YVO$_4$ in the
literature. Table \ref{tab:crystal} lists the refractive index difference of the ordinary and
extraordinary axis of YVO$_4$ and the calculated length of 
a compensation crystal for the wavelengths  $\lambda_s=894.3$\,nm and
$\lambda_i=1313.1$\,nm for different sources.

\begin{table}[h]
\centering
\caption{Reported refractive index difference $\Delta n=n_e-n_o$ of the ordinary and
extraordinary axis of YVO$_4$ from different sources and the calculated corresponding length of compensation crystal
for $\lambda_s=894.3$\,nm and $\lambda_i=1313.1$\,nm.\label{tab:crystal}}

\begin{tabular}{llll}
Source &   $\Delta n_s$ & $\Delta n_i$ &L [mm]   \\\hline
Manufacturer (Foctek) \cite{foctek_photonics_inc._yvo4_????} & 0.211014	& 0.205272 & 138.7\\
Zelmon, et al. \cite{zelmon_revisiting_2010} & 0.211408	& 0.205449&  154.0   \\
Sato, et al. \cite{sato_highly_2014} & 0.213228 & 0.207146 &  172.0   \\
Handbook of Optics \cite{bass_handbook_2009} & 0.209959 & 0.204679 &  178.8  \\
\end{tabular}
\end{table}
%

\begin{figure} 
        \centering\includegraphics[width=\columnwidth]{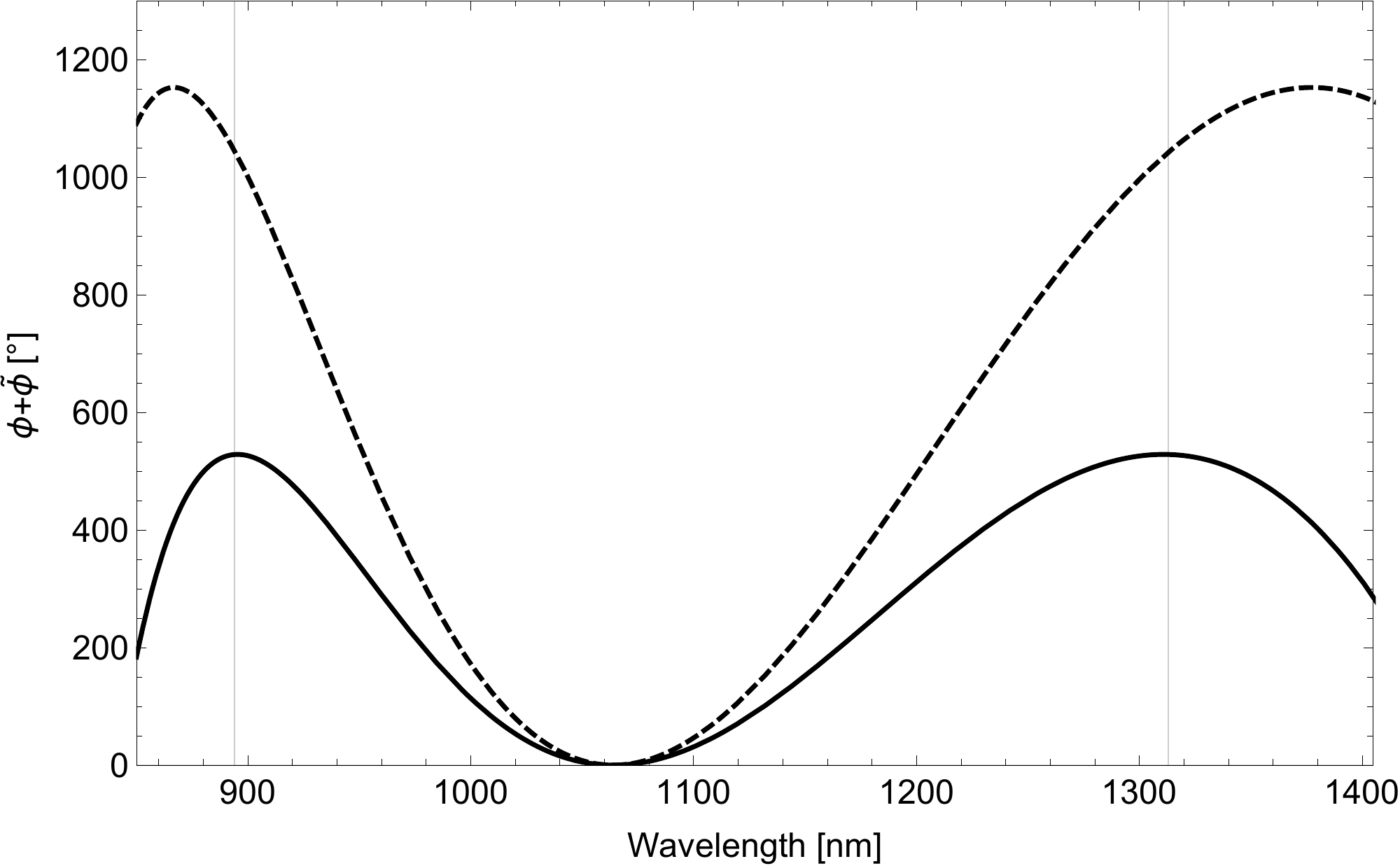}
\caption{Calculated phase $\phi+\tilde \phi$ as a function of the wavelength for two different lengths of
        the compensation crystal (constant offset subtracted). A flat phase is
		  obtained for $\tilde L=154$\,mm (\blacksolid) for the two target
		  wavelengths $\lambda_s=894.3$\,nm and $\lambda_i=1313.1$\,nm (vertical
		  lines). The plateaus of the flat phases are shifted considerably for $\tilde L
		  = 153$\,mm (\blackdashed).
                \label{fig:yvo4length}} 
\end{figure}

Figure~\ref{fig:yvo4length} shows the calculated phase $ \phi + \tilde \phi$ as
a function of the wavelength for two different lengths of the compensation
crystal. The two target wavelengths $\lambda_s=894.3$\,nm and
$\lambda_i=1313.1$\,nm are plotted as vertical lines. 
%
%

For $\pm 1$\,mm  crystal length, the two flat phase positions vary by $\pm
50$\,nm. Thus with adding or removing additional few mm slabs of YVO$_4$ the total compensation can be tuned by several hundred nanometers.

\begin{figure}
		  \centering\includegraphics[width=\columnwidth]{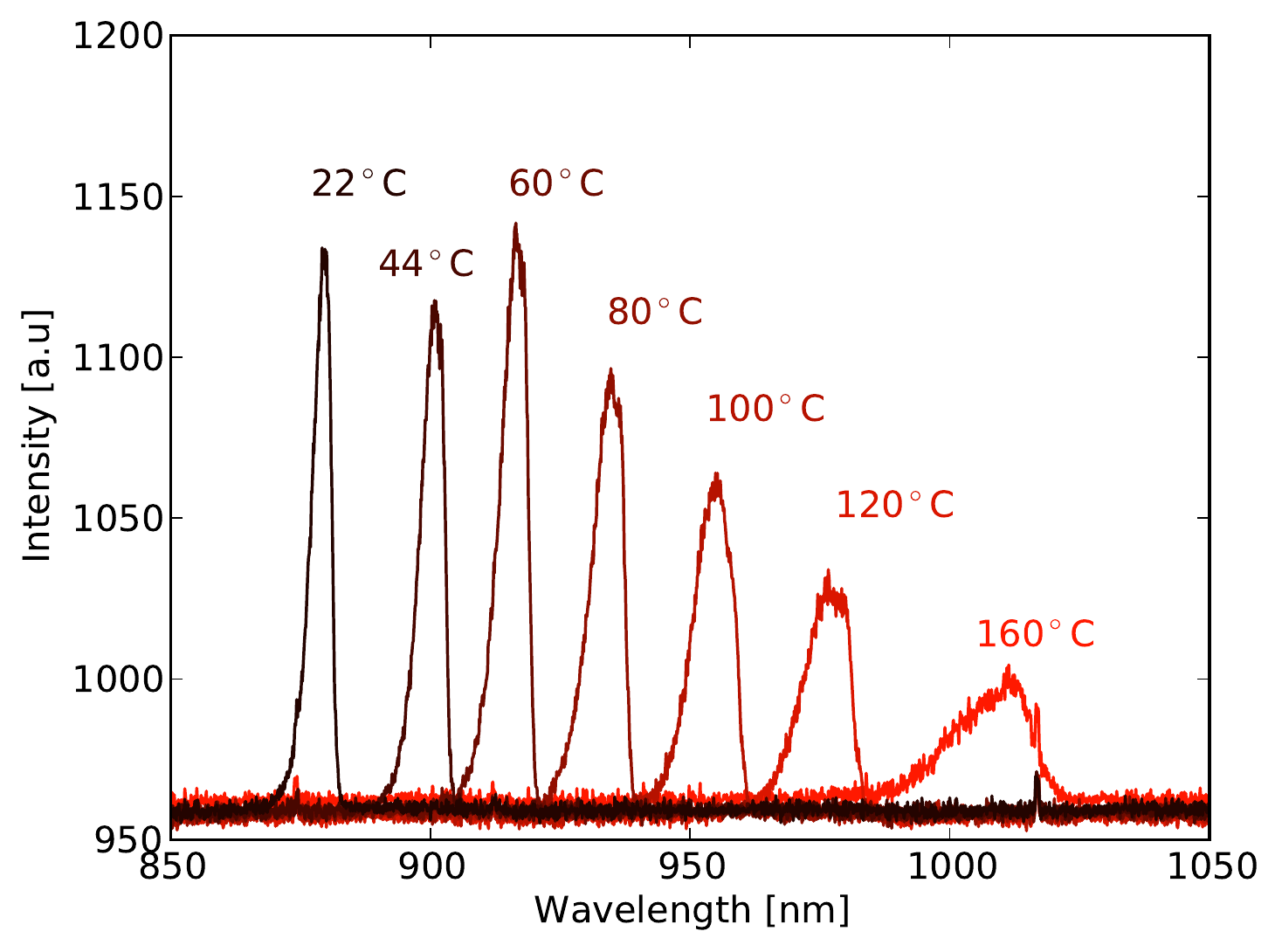} 
\caption{ 
        Spectra of signal photons for different temperatures of the non-linear crystal. 
        The decreasing intensities with higher temperatures are due to decreasing sensitivity of the 
        spectrometer CCD camera. At 1064\,nm the signal and idler photons are 
        degenerate. 
        \label{fig:T}} 
\end{figure}

Finally, Fig.~\ref{fig:T} demonstrates the tuning capability of our source. The
crystal oven of the non-linear crystal can be heated up to 160\,$^\circ$C.
The
measured tuning range of the signal photon wavelength for the chosen 7\,$\upmu$m
grating extends from 870\,nm to 1100\,nm, this is roughly 230\,nm. 
The corresponding idler wavelength spans from 1124\,nm to 1345\,nm (not shown). The measurement was performed without compensation crystal. 

\section{Determining the Optimal Crystal Length}

\begin{figure}
\centering\includegraphics[width=\columnwidth]{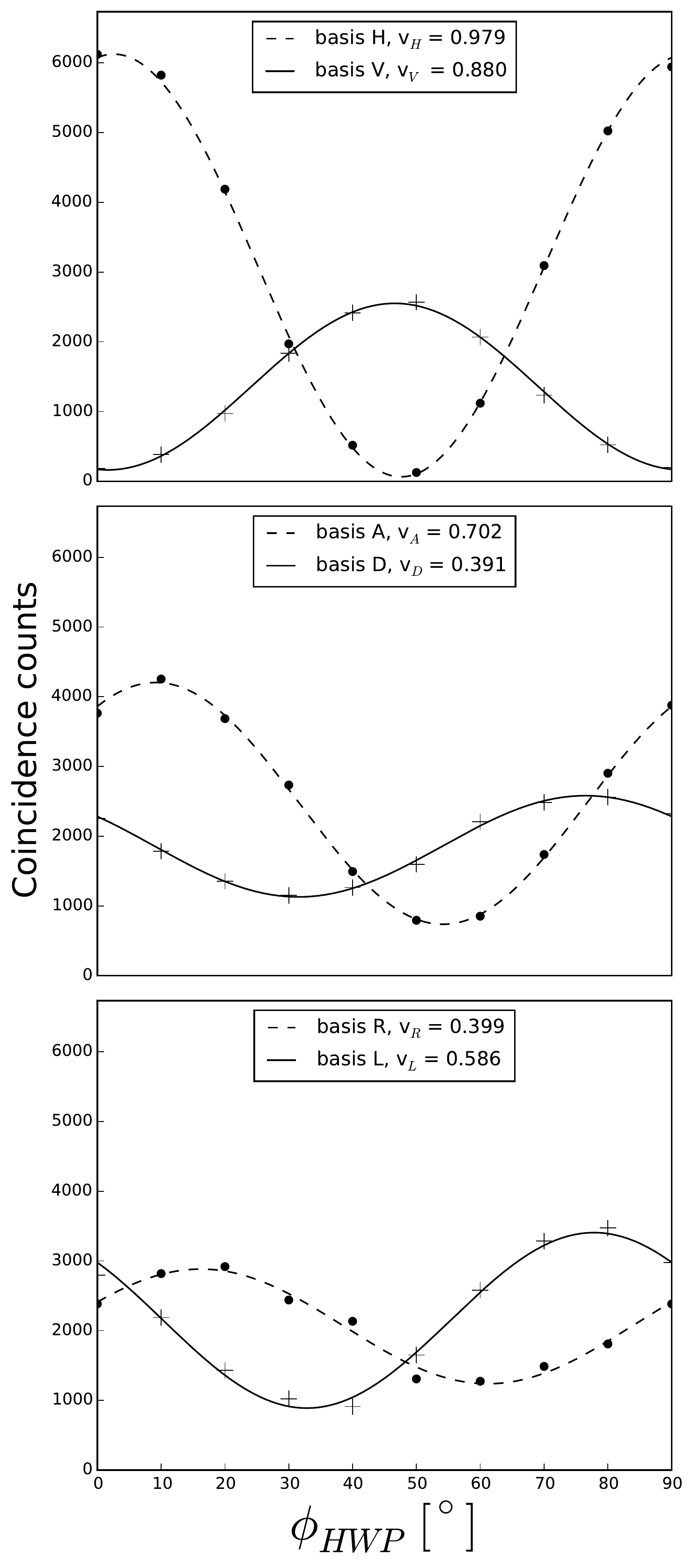}
\caption{
		  Measured coincidence counts as a function of the orientation of the
		  half-wave plate in different basis combinations. The dashed and solid
		  lines are fits to sine-functions. The resulting visibilities $v_i$ from
					 Eq.~\eqref{eq:vis} are indicated.
		  \label{fig:visibility}}
\end{figure}

The wave-plates in each arm (after DM2 in Fig.~\ref{fig:setup}) can be
rotated such that photon coincidences in different bases can be measured. Both
polarizers are fixed
to horizontal polarization, such that in each arm the basis $\ket{H}$ is measured for a
half-wave plate position of $\Theta_\text{HWP}=0$. Rotating the
half-wave plate to $\Theta_\text{HWP}=\pm22.5^\circ$ the diagonal $\ket{D}$ and
anti-diagonal $\ket{A}$ basis is measured, respectively. Adding a quarter-wave
plate the left $\ket{L}$ and right $\ket{R}$ basis can be measured.   
Each arm can be set to an individual polarization. For example, setting both arms
to left-circularly polarization is denoted
$\ket{L}_\text{signal}\ket{L}_\text{idler}=\ket{LL}$ in the following.

The procedure to find the right compensation crystal relies on measuring
polarization-sensitive coincidence counts as a function of the orientation of
the half-wave plate $\theta_{HWP}$ in one arm.
Fig.~\ref{fig:visibility} shows such measurements in different bases combinations. 
The measured curve can be fitted to a sine function and the visibility
can be derived (see next Section for details).
Then, thin
slabs of YVO$_4$ are added, until the visibility cannot be enhanced further. 
In this way an
optimum total length of the compensation crystal can be found.

Unfortunately, tuning by adding thin slabs of additional
compensation crystals does not provide sufficient accuracy. 
Therefore a fine tuning of the
phase between 0 and $\pi$ to generate a specific Bell-state is necessary.

\begin{figure} 
		  \centering\includegraphics[width=\columnwidth]{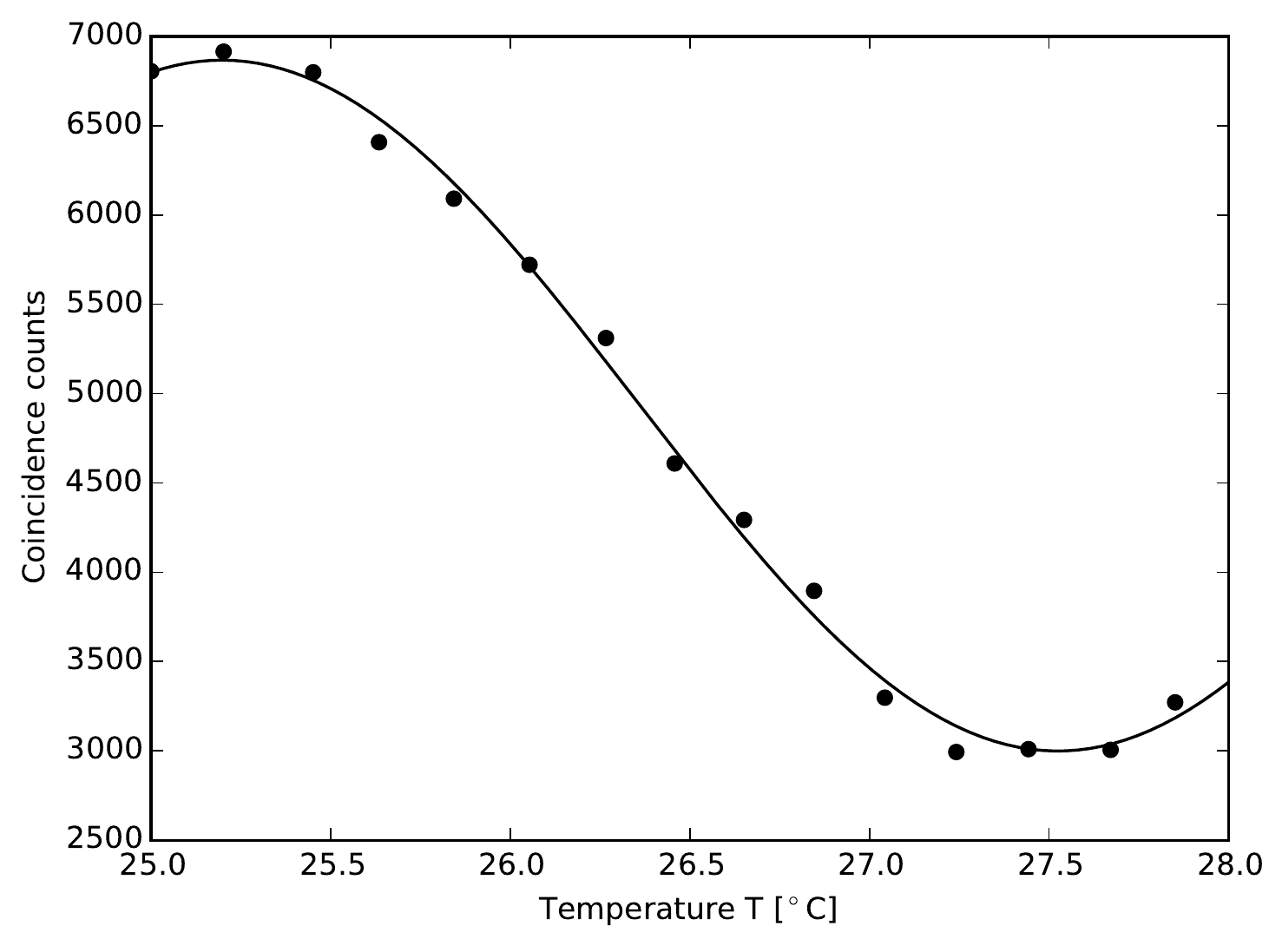}
		  \caption{Measured coincidence counts in
					 the $\ket{AA}$ basis (both polarizers anti-diagonal) as a function of the temperature
		  of a 30\,mm compensation crystal slab. The total crystal
					 length is 153\,mm.  \label{fig:shift}} 
\end{figure}
In order to do this we changed the temperature
of a  30\,mm compensation crystal slab.
Figure~\ref{fig:shift} shows the measured coincidence counts between signal and
idler photons in the $\ket{AA}$ basis as a function of the temperature of the compensation crystal. 
We find a phase shift of $\pi$, i.e., between two different Bell-states, for
$\sim2.4\,^\circ$C temperature difference.

For the wavelengths 
$\lambda_s=894.3$\,nm and $\lambda_i=1313.1$\,nm we find a optimal total crystal length of
$L=153$\,mm. The total length was composed of 7 slabs YVO$_4$ of 2\,cm length and
one slab of 1\,cm, 2\,mm and 1\,mm length. The total length is close to some of the reported data on the
refractive index of YVO$_4$ in literature
\cite{zelmon_revisiting_2010}, even though they investigated 0.5\% Nd-doped YVO$_4$.
However, it deviates from the value reported by other studies
\cite{sato_highly_2014},
as well as from the manufacturer specification (Foctek Inc.)
\cite{foctek_photonics_inc._yvo4_????}.

With the experimentally determined crystal parameters, it is in
principle straightforward to estimate the compensation crystal length also for other pairs of wavelengths. Additional crystal slabs can be added or removed allowing for a wide tuning range.

\section{Verifying Entanglement}
To quantify the degree of entanglement for optimized conditions, we measured the
$\ket{\phi^+}$ Bell-state fidelity  $F_{\phi^+}$ of the created state $\rho$,
\begin{equation}
		  F_{\phi^+}=\bra{\phi^+}\rho\ket{\phi^+}.
\label{eq:bell}
\end{equation}
A fidelity $F_{\phi^+}>\frac{1}{2}$ indicates entanglement
\cite{white_measuring_2007}.
In order to relate the fidelity to the coincidence probabilities we use 
the two-photon polarization basis states  $\ket{u_1}=\ket{HH}$,  $\ket{u_2}= \ket{HV}$,  $\ket{u_3}= \ket{VH}$,  $\ket{u_4}= \ket{VV}$. After expanding $\rho$ in this basis and using 
\begin{equation}
		  \ket{\phi^+}=\frac{1}{\sqrt{2}}\left(\ket{u_1}+\ket{u_4}\right),
\end{equation} 
we get 
\begin{align}
		  F_{\phi^+}&=\sum_{k,l=1}^4
					 \rho_{kl}\bral{\phi^+}\ket{u_k}\bral{u_l}\ket{\phi^+}\nonumber\\
 &= \frac{1}{2}\left( \rho_{11}+\rho_{14}+\rho_{41}+\rho_{44} \right),
\end{align}
where $\rho_{kl}=\bra{u_k}\rho\ket{u_l}$. 
The diagonal elements $\rho_{11}$ and $\rho_{44}$ are the probabilities $P_H$ and $P_V$ of finding a photon pair in the
states $\ket{HH}$ and  $\ket{VV}$, respectively. 
The off-diagonal elements can be expressed with the help of the diagonal and
circular basis, 
\begin{align}
		  P_D+P_A &= \bra{DD}\rho\ket{DD}+\bra{AA}\rho\ket{AA}\nonumber\\
		  &= \frac{1}{2}\left( 1+\rho_{14}+\rho_{41}+\rho_{23}+\rho_{32} \right),\\
		  P_R+P_L &= \frac{1}{2}\left( 1-\rho_{14}-\rho_{41}+\rho_{23}+\rho_{32} \right)
\end{align}
and thus 
\begin{align}
		  P_D+P_A-\left(P_R+P_L\right)=\rho_{14}+\rho_{41}.
		  \label{}
\end{align}
We therefore obtain the fidelity as a function of the probabilities in the three different
bases, 
\begin{equation}
        F_{\phi^+}=\frac 1 2 \left [ P_D+P_A - \left( P_R + P_L \right) + P_H + P_V \right ].
        \label{eq:fprop}
\end{equation}
These probabilities can be measured as they are connected to the coincidence count rates $N$,
\begin{equation}
        P_i =\frac{N_{ii}}{N_{ii}+N_{ij}+N_{jj}+N_{ji}}= \frac{N_{ii}}{N_{tot}},
 \end{equation}
where  $i$ refers to  the one photon polarization state orthogonal to $j$, that is $i/j=H/V, D/A$, or $L/R$.
After introducing 
\begin{equation}
		  V_{ij}=\frac{N_{ii}+N_{jj}-(N_{ij}+N_{ji})}{N_{ii}+N_{jj}+N_{ij}+N_{ji}}=\frac{2\left( N_{ii}+N_{jj} \right)}{N_{tot}}-1
		  \label{}
\end{equation}
Eq. (\eqref{eq:fprop}) yields the alternative representation \cite{steinlechner_phase-stable_2013}
\begin{equation}
 F_{\phi^+}=\frac{1}{4}\left( 1+V_{HV}+V_{DA}-V_{LR} \right).
  \label{eq:fvis}
\end{equation}

To determine the fidelity, we measured the visibilities 
\begin{equation}
		  v_i = \frac{N_{ii}-N_{ij}}{N_{ii}+N_{ij}}=\frac{A_i}{C_i}
		  \label{eq:vis}
\end{equation} in the 3 different bases, as
shown in Fig.~\ref{fig:visibility}.
In the measurement the basis in one arm
is fixed, while the basis in the other arm is rotated between $i$, and its orthogonal counter part $j$.
The visibilities are then fitted with a sine function
$A_i\sin(4 \phi_{HWP}+\text{const.})+C_i$.
Since $C_i = (N_{ii}+N_{ij})/2$ and $A_i = (N_{ii}-N_{ij})/2$, we have
\begin{equation}
		  V_{ij}=\frac{A_i+A_j}{C_i+C_j}\,.
		  \label{eq:fit}
\end{equation}
With this we find a fidelity of $F_{\phi^+}=0.753 \pm 0.021$ for our source at a photon pair
generation rate of 5.8\,Mcps/mW and spectral linewidths of 560\,GHz, which
corresponds to
1.5\,nm and 3.3\,nm at  
$\lambda_s=894.3$\,nm and $\lambda_i=1313.1$\,nm, respectively. 
\hyphenation{extra-ordinarily}

The limited visibility can be explained by temperature fluctuations of the
compensation crystals. The total length of the compensation crystal is
$L=153$\,mm, but only a small fraction (3\,cm) of the crystal is temperature
stabilized inside the crystal oven. 
Small changes in temperature modify the optical path-lengths of extraordinarily
and ordinarily polarized photons and lead to a fluctuating phase. For our
configuration the temperature fluctuations at the compensation crystals of $\pm
1$ K are the strongest factor that reduces the fidelity (see also \cite{steinlechner_phase-stable_2013}). 
With an improved temperature stabilization and a thermally isolated housing of the compensation crystals fidelities above 
		  95$\%$ should be possible \cite{steinlechner_phase-stable_2013}.

\section{Conclusion}
We demonstrated a novel folded-sandwich scheme for the generation of two-color
entangled photons which uses a Fresnel rhomb as a geometrical quarter-wave
plate. By this, all optical components are more easily adapted to wide
combinations of wavelengths. For example, no three-color beam splitters as in a
Sagnac configuration are required. Adjusting the compensation crystal
length offers a tuning capability over more than 100\,nm.
Our source is a viable tool to provide highly non-degenerate entangled photons for quantum hybrid architectures, in particular when solid-state emitters with an a priori unpredictable transition frequency are involved.

\section{Acknowledgements}
This work was funded by BMBF (Q.com-H).
Funding by DFG through SFB 787 is acknowledged by O.D. O.D.~likes to thank Sven
Ramelow, Fabian Steinlechner and Amir Moqanaki for their hospitality during his
stay in Vienna and their continuous input and support on the experimental design.
This work was also supported by project EMPIR 14IND05 MIQC2 (the EMPIR
initiative is co-funded by the European Union's Horizon 2020 research and
innovation programme and the EMPIR Participating States).

\bibliographystyle{spphys}
\bibliography{biblio}

\begin{thebibliography}{10}
\providecommand{\url}[1]{{#1}}
\providecommand{\urlprefix}{URL }
\expandafter\ifx\csname urlstyle\endcsname\relax
  \providecommand{\doi}[1]{DOI \discretionary{}{}{}#1}\else
  \providecommand{\doi}{DOI \discretionary{}{}{}\begingroup
  \urlstyle{rm}\Url}\fi

\bibitem{briegel_quantum_1998}
H.J. Briegel, W.~D{\"u}r, J.I. Cirac, P.~Zoller, Phys. Rev. Lett.
  \textbf{81}(26), 5932 (1998).
\newblock \doi{10.1103/PhysRevLett.81.5932}.
\newblock \urlprefix\url{http://link.aps.org/doi/10.1103/PhysRevLett.81.5932}

\bibitem{volz_observation_2006}
J.~Volz, M.~Weber, D.~Schlenk, W.~Rosenfeld, J.~Vrana, K.~Saucke,
  C.~Kurtsiefer, H.~Weinfurter, Phys. Rev. Lett. \textbf{96}(3), 030404 (2006).
\newblock \doi{10.1103/PhysRevLett.96.030404}.
\newblock \urlprefix\url{http://link.aps.org/doi/10.1103/PhysRevLett.96.030404}

\bibitem{blinov_observation_2004}
B.B. Blinov, D.L. Moehring, L.M. Duan, C.~Monroe, Nature \textbf{428}(6979),
  153 (2004).
\newblock \doi{10.1038/nature02377}.
\newblock
  \urlprefix\url{http://www.nature.com/nature/journal/v428/n6979/full/nature02377.html}

\bibitem{gao_observation_2012}
W.B. Gao, P.~Fallahi, E.~Togan, J.~Miguel-Sanchez, a.~Imamoglu, Nature
  \textbf{491}(7424), 426 (2012).
\newblock \doi{10.1038/nature11573}.
\newblock \urlprefix\url{http://www.nature.com/doifinder/10.1038/nature11573}

\bibitem{togan_quantum_2010}
E.~Togan, Y.~Chu, A.S. Trifonov, L.~Jiang, J.~Maze, L.~Childress, M.V.G. Dutt,
  A.S. Sørensen, P.R. Hemmer, A.S. Zibrov, M.D. Lukin, Nature
  \textbf{466}(7307), 730 (2010).
\newblock \doi{10.1038/nature09256}.
\newblock
  \urlprefix\url{http://www.nature.com/nature/journal/v466/n7307/full/nature09256.html}

\bibitem{wallraff_strong_2004}
A.~Wallraff, D.I. Schuster, A.~Blais, L.~Frunzio, R.S. Huang, J.~Majer,
  S.~Kumar, S.M. Girvin, R.J. Schoelkopf, Nature \textbf{431}(7005), 162
  (2004).
\newblock \doi{10.1038/nature02851}.
\newblock
  \urlprefix\url{http://www.nature.com/nature/journal/v431/n7005/full/nature02851.html}

\bibitem{hofmann_heralded_2012}
J.~Hofmann, M.~Krug, N.~Ortegel, L.~G{\'e}rard, M.~Weber, W.~Rosenfeld,
  H.~Weinfurter, Science \textbf{337}(6090), 72 (2012).
\newblock \doi{10.1126/science.1221856}.
\newblock \urlprefix\url{http://www.sciencemag.org/content/337/6090/72}

\bibitem{bernien_heralded_2013}
H.~Bernien, B.~Hensen, W.~Pfaff, G.~Koolstra, M.S. Blok, L.~Robledo, T.H.
  Taminiau, M.~Markham, D.J. Twitchen, L.~Childress, R.~Hanson, Nature
  \textbf{497}(7447), 86 (2013).
\newblock \doi{10.1038/nature12016}.
\newblock
  \urlprefix\url{http://www.nature.com/nature/journal/v497/n7447/abs/nature12016.html}

\bibitem{kwiat_new_1995}
P.G. Kwiat, K.~Mattle, H.~Weinfurter, A.~Zeilinger, A.V. Sergienko, Y.~Shih,
  Phys. Rev. Lett. \textbf{75}(24), 4337 (1995).
\newblock \doi{10.1103/PhysRevLett.75.4337}.
\newblock \urlprefix\url{http://link.aps.org/doi/10.1103/PhysRevLett.75.4337}

\bibitem{pelton_bright_2004}
M.~Pelton, P.~Marsden, D.~Ljunggren, M.~Tengner, A.~Karlsson, A.~Fragemann,
  C.~Canalias, F.~Laurell, Optics Express \textbf{12}(15), 3573 (2004).
\newblock \doi{10.1364/OPEX.12.003573}.
\newblock
  \urlprefix\url{http://www.opticsexpress.org/abstract.cfm?URI=OPEX-12-15-3573}

\bibitem{hentschel_three-color_2009}
M.~Hentschel, H.~H{\"u}bel, A.~Poppe, A.~Zeilinger, Optics express
  \textbf{17}(25), 23153 (2009).
\newblock
  \urlprefix\url{http://www.opticsinfobase.org/abstract.cfm?uri=oe-17-25-23153}

\bibitem{stuart_flexible_2013}
T.E. Stuart, J.A. Slater, F.~Bussi{\`e}res, W.~Tittel, Physical Review A
  \textbf{88}(1) (2013).
\newblock \doi{10.1103/PhysRevA.88.012301}.
\newblock \urlprefix\url{http://link.aps.org/doi/10.1103/PhysRevA.88.012301}

\bibitem{trojek_collinear_2008}
P.~Trojek, H.~Weinfurter, Applied Physics Letters \textbf{92}(21), 211103
  (2008).
\newblock \doi{10.1063/1.2924280}.
\newblock
  \urlprefix\url{http://scitation.aip.org/content/aip/journal/apl/92/21/10.1063/1.2924280}

\bibitem{steinlechner_phase-stable_2013}
F.~Steinlechner, S.~Ramelow, M.~Jofre, M.~Gilaberte, T.~Jennewein, M.W.
  Mitchell, V.~Pruneri, J.P. Torres, Optics Express \textbf{21}(10), 11943
  (2013).
\newblock \doi{10.1364/OE.21.011943}.
\newblock
  \urlprefix\url{http://www.opticsinfobase.org/abstract.cfm?URI=oe-21-10-11943}

\bibitem{ding_tuning_2010}
F.~Ding, R.~Singh, J.D. Plumhof, T.~Zander, V.~Kř{\'a}pek, Y.H. Chen,
  M.~Benyoucef, V.~Zwiller, K.~D{\"o}rr, G.~Bester, A.~Rastelli, O.G. Schmidt,
  Phys. Rev. Lett. \textbf{104}(6), 067405 (2010).
\newblock \doi{10.1103/PhysRevLett.104.067405}.
\newblock
  \urlprefix\url{http://link.aps.org/doi/10.1103/PhysRevLett.104.067405}

\bibitem{akopian_hybrid_2011}
N.~Akopian, L.~Wang, A.~Rastelli, O.G. Schmidt, V.~Zwiller, Nat Photon
  \textbf{5}(4), 230 (2011).
\newblock \doi{10.1038/nphoton.2011.16}.
\newblock
  \urlprefix\url{http://www.nature.com/nphoton/journal/v5/n4/full/nphoton.2011.16.html}

\bibitem{siyushev_molecular_2014}
P.~Siyushev, G.~Stein, J.~Wrachtrup, I.~Gerhardt, Nature \textbf{509}(7498), 66
  (2014).
\newblock \doi{10.1038/nature13191}.
\newblock
  \urlprefix\url{http://www.nature.com/nature/journal/v509/n7498/full/nature13191.html}

\bibitem{alexander_franzen_componentlibrary:_????}
{Alexander Franzen}.
\newblock {ComponentLibrary}: a free vector graphics library for optics.
  {Licensed} under a {Creative} {Commons} {Attribution}-{NonCommercial} 3.0
  {Unported} {License}.
\newblock \urlprefix\url{http://www.gwoptics.org/ComponentLibrary/}

\bibitem{foctek_photonics_inc._yvo4_????}
{Foctek Photonics, Inc.}
\newblock {YVO}4.
\newblock \urlprefix\url{http://www.foctek.net/products/YVO4.htm}

\bibitem{zelmon_revisiting_2010}
D.E. Zelmon, J.M. Northridge, D.~Perlov, Appl. Opt. \textbf{49}(4), 644 (2010).
\newblock \doi{10.1364/AO.49.000644}.
\newblock \urlprefix\url{http://ao.osa.org/abstract.cfm?URI=ao-49-4-644}

\bibitem{sato_highly_2014}
Y.~Sato, T.~Taira, Opt. Mater. Express \textbf{4}(5), 876 (2014).
\newblock \doi{10.1364/OME.4.000876}.
\newblock
  \urlprefix\url{http://www.opticsinfobase.org/ome/abstract.cfm?URI=ome-4-5-876}

\bibitem{bass_handbook_2009}
M.~Bass, C.~DeCusatis, J.~Enoch, V.~Lakshminarayanan, G.~Li, C.~MacDonald,
  V.~Mahajan, E.V. Stryland, \emph{Handbook of {Optics}, {Third} {Edition}
  {Volume} {IV}: {Optical} {Properties} of {Materials}, {Nonlinear} {Optics},
  {Quantum} {Optics} (set): {Optical} {Properties} of {Materials}, {Nonlinear}
  {Optics}, {Quantum} {Optics} (set)} (McGraw Hill Professional, 2009)

\bibitem{white_measuring_2007}
A.G. White, A.~Gilchrist, G.J. Pryde, J.L. O'Brien, M.J. Bremner, N.K.
  Langford, JOSA B \textbf{24}(2), 172 (2007).
\newblock
  \urlprefix\url{http://www.opticsinfobase.org/abstract.cfm?uri=josab-24-2-172}

\end{thebibliography}


\end{document}